# Magneto-optical Kerr spectra of gold induced by spin accumulation

Víctor H. Ortiz, Sinisa Coh and Richard B. Wilson

*Department of Mechanical Engineering and Materials Science and Engineering Program, University of California, Riverside, California 92521, USA*

**Abstract.** We report the magneto optic Kerr effect (MOKE) angle of Au magnetically excited by spin accumulation. We perform time resolved polar MOKE measurements on Au/Co heterostructures. In our experiment, the ultrafast optical excitation of the Co drives spin accumulation into an adjacent Au layer. The spin accumulation, together with spin-orbit coupling, leads to non-zero terms in the off-diagonal conductivity tensor of Au, which we measure by recording the polarization and ellipticity of light reflected from the Au surface for photon energies between 1.3 and 3.1 eV. At energies near the interband transition threshold of Au, the Kerr rotation per A/m exceeds 1 µrad. Typical values for Kerr rotation per moment in transition ferromagnetic metals like Ni are < 10 nrad per A/m, while predicted values for heavy metals like Pt or W are < 13 nrad per A/m. The exceptional sensitivity of the optical properties of Au to spin magnetic moments make Au to be an exceptionally sensitive optical magnetometer, with potential applications in the development of optospintronic technologies.

## I. Introduction

Spin-orbit interactions affect how the electrons of a metal move in response to electromagnetic fields[1]. The effect of spin-orbit interactions on motion is opposite for up vs. down electrons. In magnetic metals, where the number density of up vs. down spins is different, spin-orbit interactions give rise to the well-known magneto-optic Kerr and Faraday effects[2]. These magneto-optic effects describe how the polarization of reflected or transmitted electromagnetic waves differs from the incident wave. Alternatively, in non-magnetic metals, the net effect spin-orbit interactions on optical properties is zero. This is because, in their ground state, non-magnetic metals have equal number densities of up vs. down spins.

In the presence of an external magnetic field, non-magnetic metals display magneto-optic effects for two reasons. First, the static magnetic field applies a Lorentz force, which affects the motion of the electrons and gives rise to non-zero off diagonal terms in the optical conductivity tensor[3]. Second, applying a magnetic field to a metal leads to Zeeman splitting, i.e. leads to spin and orbital magnetic moments[4]. The direct effect of the Lorentz force exerted by the static magnetic field on the optical properties of a metal is analogous to the Hall effect[3], while the combined effect of spin-accumulation and spin-orbit interactions on the optical properties is analogous to the anomalous Hall effect[5]. Numerous experimental and theoretical studies have documented the magneto-optical properties of non-magnetic metals in the presence of a static magnetic field[4], [6], [7]. Alternatively, the physics of how only spin-accumulation induces magneto-optical properties in the absence of orbital moments or Lorentz fields is not well understood. Prior experimental studies of spin-accumulation induced MOKE in non-magnetic metals were limited to a single photon energy[5], [8]. No prior experimental studies of spin-accumulation induced MOKE have reported spectra across a wide energy range.

Understanding how only spin accumulation and spin-orbit interactions affect the optical properties of non-magnetic metals is important for a number of emerging fields. Magneto-optic Kerr effect (MOKE) signals from non-magnetic metals allow experimentalists to measure a number of spin-transport material properties. Stamm et al. used MOKE signals to measure the spin-Hall conductivity and spin-diffusion length of Pt and W[9]. Melnikov et al. used time-resolved measurements of non-linear magneto-optic properties to characterize spin-polarized hot carrier transport in Fe/Au bilayers[8]. Time-resolved MOKE signals from non-magnetic metals allows for the measurement of spin caloritronic transport properties of adjacent magnetic materials[5], [10]–[12]. Hofherr et al. used time-resolved measurements of the complex Kerr effect from Au/Ni bilayers to separately measure the ultrafast magnetization dynamics of both layers[13].

Magneto-plasmonics is another field of study where understanding how spin-orbit interactions affect optical properties is critical[14]–[22]. The ability of plasmonic resonances to enhance magneto-optical effects such as the inverse Faraday has received significant recent attention[18], [19], [22]–[25]. The inverse Faraday effect describes the ability of circularly polarized light to induce a magnetic moment in a metal[23], [24]. Interpretation of MOKE signals that arise from the inverse Faraday effect is challenging, because the magnetic moment induced by IFE can include both orbital and spin contributions[26].

In this work, we report experimental measurements of the complex Kerr spectra of Au in the presence of spin accumulation. We perform a series of time-resolved magneto-optic Kerr effect (TR-MOKE) measurements of a Au/Co multilayer as a function of laser energy. Our experiments span photon energies between 1.3 and 3.1 eV. We observe a resonance-like behavior in the Kerr spectra of Au near an interband transition threshold energy of 2.5 eV. The Kerr angle of Au is ~20x larger at energies near the interband transition threshold then in the near infrared. On a per magnetic moment basis, we find that the Kerr angle of Au near 2.5 eV is ~100x greater than is typical in ferromagnetic metals like Ni. To our knowledge, the only material with a comparable magneto-optical response per moment is antiferromagnetic metal $Mn_3Sn$, whose large Kerr effect origin relates to ferroic ordering of magnetic octupoles[27].

We also performed density functional theory predictions the MOKE spectra of gold. Our theory includes the intrinsic effects of spin-orbit interactions on the band-structure, but neglects spin-orbit scattering effects like side-jump scattering and skew-scattering. We also neglect the effect of spin-orbit interactions and surfaces on the band-structure of Au[28]. Overall, the agreement between theory and experiment is excellent, suggesting intrinsic band-structure effects adequately explain the magnitude of observed MOKE signals.

## II. Results

### A. Theory

We use the first principles calculated electron band structure energies $E_{n\bm{k}}$ and orbitals $\psi_{n\bm{k}}$ within the perturbative approach to evaluate the optical conductivity of Au[29], [30],

$$\sigma_{\alpha\beta}(\omega) = \frac{i\,e^2\hbar}{(2\pi)^3} \lim_{\bm{q}\to 0} \int d\bm{k} \sum_{n,m} \frac{f_{m\bm{k}+\bm{q}} - f_{n\bm{k}}}{E_{m\bm{k}+\bm{q}} - E_{n\bm{k}}} \frac{\langle\psi_{n\bm{k}}|v_\alpha|\psi_{m\bm{k}+\bm{q}}\rangle \langle\psi_{m\bm{k}+\bm{q}}|v_\beta|\psi_{n\bm{k}}\rangle}{E_{m\bm{k}+\bm{q}} - E_{n\bm{k}} - \hbar\omega - i\eta_{mn\bm{k}}/2}, \quad (1)$$

The conductivity $\sigma_{\alpha\beta}$ describes the current in direction $\alpha$ in response to electric field pointing in direction $\beta$. Eq. (1) is a summation over possible electronic transitions between states in band $m$ at wavevector $\bm{k} + \bm{q}$ to states in band $n$ and wavevector $\bm{k}$. The limit $\bm{q} \to 0$ indicates that we include here both intraband and interband contributions to the optical conductivity. The Fermi-Dirac distribution occupation factor is denoted as $f$ while velocity operator is $v$. $\eta_{mn\bm{k}}$ describes the effect of electronic scattering rates on transitions due to electron scattering. We model $\eta_{mn\bm{k}}$ with spin-independent electron-electron and electron-phonon scattering which reproduces both first-principles calculations[31] as well as experimental data for the imaginary part of the diagonal component of the conductivity tensor.

We calculate electron wavefunctions and energies with the PBEsol+U approach as implemented in the Quantum ESPRESSO package[32]. We use U = 2.7 eV following Brown et al[33]. We sample the charge density on a 12×12×12 mesh of k-points in the equivalent one-atom unit cell. We use Wannier interpolation to converge the optical conductivity on a 300×300×300 mesh of k-points.

From the computed diagonal and off-diagonal optical conductivity, we calculate the complex Kerr angle. In the polar MOKE configuration ($\hat{z}$ perpendicular to the sample plane), the complex Kerr angle is

$$\theta_k(\omega) + i\varepsilon_k(\omega) = \frac{\sigma_{yx}(\omega)}{\sigma_{xx}(\omega)\sqrt{1+i\frac{\sigma_{xx}(\omega)}{\omega\epsilon_0}}}. \tag{2}$$

Here, $\theta_k$ is the Kerr rotation and $\varepsilon_k$ is the ellipticity. $\theta_k$ describes rotation of the major axis of polarization and is caused by differences in the indices of refraction for left vs. right circularly polarized light. $\varepsilon_k$ describes a change in the ellipticity and is caused by differences in absorption for left vs. right circularly polarized light.

### B. Time-resolved magneto-optical Kerr effect experiment

To measure the magneto-optical spectra of Au, we use ultrafast demagnetization of an adjacent ferromagnetic layer to temporarily magnetize the Au film without any external magnetic field. In Fig. 1a we depict the schematic of the system we used for the measurement. We utilize a Au/Co multilayer system as the ferromagnetic layer with perpendicular magnetic anisotropy (PMA)(Fig. 1b)[34]. The sample geometry is $d_{Au}$ Au/ [1nm-Co/2nm-Au]$_{x4}$ /2nm-A /4nm-Ta/Sapphire. The top gold film is a wedge layer, with varied thickness $d_{Au}$ between 0 to 300 nm. In our pump/probe experiments, the pump beam causes an increase on the temperature of the Au/Co ferromagnetic multilayer, resulting in ultrafast demagnetization of the Au/Co multilayer[35]. Ultrafast demagnetization of the Au/Co multilayer leads to an injection of spin current in adjacent non-magnetic layers[5], [10]. Spin accumulation in the Au layer causes a rotation of the polarization state of a time-delayed reflected probe beam.

Following pump heating of the Au/Co multilayer, we observe transient changes in the Au reflectance. In Fig. 2, we show TDTR data taken with pump and probe energy of $E$ = 2.60 eV.

(All the TDTR time-delay data is shifted so the maximum change in the in-phase signal corresponds to t = 0, this time shift is used in the TRMOKE data analysis later.) The TDTR data has distinct dynamics on timescales of picoseconds, hundreds of picoseconds, and nanoseconds. The initial rise in thermoreflectance from 0 to 1 ps is due to thermal transport by nonequilibrium electrons[5]. The subsequent decay from 1-3 ps is due to electron-phonon thermalization. Then, the thermoreflectance signal increases for ~500 ps as heat diffuses across the thick Au layer. On timescales longer the 0.5 ns, the thermoreflectance signal decays as heat diffuses out of the metal multilayer and into the sapphire substrate.

We use picosecond acoustic signals in our TDTR data to determine the Au film thickness in the region where the laser is focused on the sample. Pump heating of the Au/Co multilayer generates a longitudinal acoustic pulse. After traversing the Au layer, the acoustic pulse reaches the surface of the Au layer and modifies the reflectance of the Au film. For the region of the sample where the data in Fig 2b was collected, the acoustic pulse takes ~61 ps to traverse the metal multilayer. The longitudinal speed of sound in Au is 3.45 nm/ps, corresponding to $d_{Au} \approx 210$ nm.

The primary aim of our study is to use the strength of TRMOKE signal as a function of probe wavelength to determine the magneto-optical spectra of Au. However, the wavelength dependence of the optical spectra is not the only parameter that affects signal amplitude vs. wavelength. TDTR signal amplitude also depends on absorbance of the sample, transmissivity of the objective lens, and the amplitude of the modulation of the pump beam by the electro-optical modulator. TRMOKE signal also depends on all of these factors as well as laser pulse duration. Therefore, to facilitate comparison between our measurements at different wavelengths, we normalize the raw data to account for the wavelength dependence of all these experimental parameters. In figures 2 and 3, we report normalized pump/probe data. In Fig.

2c, we compare the energy-dependence of the thermoreflectance signals in our experiments to prior measurements for Au thermoreflectance spectra. We followed the same analytical procedure as outlined in Wilson et al[36], with minor modifications to account for differences in detection electronics for our pump/probe system[37]. The magnitude and energy dependence change of sign of the thermoreflectance at E = 2.50 eV, in agreement with previous experiments[36], [38].

We now turn our attention to the results of the TR-MOKE measurements. Before performing wavelength dependent TR-MOKE studies, we conducted TR-MOKE measurements as a function of Au thickness. We did this to evaluate the experimental tradeoffs related to Au film thickness. In agreement with prior work by Choi et al.[5], we observe that the signal strength decreases with increasing Au thickness, see Fig. 3a. The decrease in signal with increasing thickness is due to spin accumulation being distributed across a larger volume as the thickness of the gold layer increases, and the finite spin-diffusion length of Au. The thickness dependence of our MOKE signals is consistnet with a Au spin-diffuion length of ~70 ± 10 nm (Fig. 3b). The decrease in MOKE signal for increasing Au thicknesses is undesirable for our purposes. But there are other advantages to a thick Au layer. For Au thicknesses not sufficiently thick, experimental signals become sensitive to the magneto optical response of the buried Co layers[5]. Since our goal is to measure the MOKE of Au, we need to eliminate any contributions to our signal from Co. Another problem we observe when the Au film is too thin is leaked pump light. For experiments with a frequency-doubled probe beam, we use optical filters to prevent pump light from reaching the photodetector. However, when conducting experiments with a near infrared probe, the probe and pump are the same energy, and so we cannot use an optical filter to prevent pump light from reaching the detector. After extensive trial and error, we discovered that a gold layer thickness of 200 nm was sufficient to prevent pump light from reaching the detector, and to remove any contribution to the MOKE signal from Co at all wavelengths. Consequently, we conducted all wavelength dependent pump/probe measurements at a region of the sample where $d_{Au} \approx 210$ nm.

To determine the amount of spin-accumulation in our samples, we use the value of 43 nrad per A/m[5], [10] for Au at 780 nm. This value for the real Kerr angle of Au per magnetic moment was arrived at via careful comparison of two types of experiments[10]. The accuracy of this value for the Kerr rotation at 780 nm is estimated to be ~30%. (Choi reports a value of ~24 nrad per A/m[5], but follow-up work reported a calibration error that led to a factor of 1.8 underestimation of the Kerr rotation[10]). For an incident pump power of 20 mW with pump and probe wavelengths of 1.6 eV, we observe a peak spin-accumulation of 2.6 A/m at a delay time of ~0.5 ps.

The results of the TR-MOKE measurements as a function of probe energy are shown in Figs. 3 and 4. This data is the primary result of our study. We performed two TR-MOKE scans at each probe energy. One scan to measure the real Kerr angle of Au, and one scan to measure the imaginary Kerr angle. All together, we performed 32 experiments at 16 probe energies. We show TR-MOKE scans in Fig. 3b at probe energies of 1.6, 2.4, 2.6, and 3.1 eV. Figure 4 reports the maximum of the transient Kerr signals as a function of probe energy. Figure 4a shows the real part of the Kerr angle (rotation), and 4b shows the imaginary part (ellipticity). In addition to the data points shown in Fig. 4, we performed several measurements at probe energies between 1.2 and 1.4 eV. However, the MOKE signal at these energies dropped below the noise floor level of our experimental technique, so at energies between 1.2 and 1.4 eV, our experiments can only bound the real and imaginary Kerr angles to be less than 10 nrad per A/m.

We compare our experimental data to the results of the density functional theory calculations in Fig. 4. DFT predicts the same spectral behavior that we observe experimentally, with maxima in the Kerr response of Au near the interband transition threshold energy of 2.5 eV. Consistent with our experimental observations, at near infrared energies below 1.7 eV, DFT predicts a real Kerr angle ~10x larger than the imaginary Kerr angle.

## III. Discussion

The most striking result of our study is our observation that the Kerr spectra for Au has a sharp Lorentzian line-shape like the kind associated with a high Q-factor oscillator. The Lorentzian is centered at the interband transition threshold of 2.5 eV with a large amplitude. The imaginary Kerr angle peaks near 2.5 eV at a value above ~ 1 μrad per A/m. The real Kerr angle peaks near 2.6 eV at ~ 1 μrad per A/m. Near the interband transition threshold, the Kerr response of Au is ~20x larger than it is at near-infrared photon energies, e.g. the 1.6 eV where the vast majority of TR-MOKE experiments are conducted. For comparison, the Kerr angle of nickel has a maximum value of ≈ 4 mrad at a photon energy of ≈ 4 eV[39]. Nickel has a magnetic moment of ≈ $5.1 \times 10^5$ A/m, so the Kerr angle per magnetization in Ni has a maximum of only 8 nrad per A/m, more than two orders of magnitude smaller than the maxima we observe for Au. The sensitivity of the optical properties of Au to spin-accumulation is also much larger than DFT predictions for other non-magnetic heavy metals. Stamm et al. report DFT predictions for the longitudinal MOKE spectra of Pt and W; using their calculated values gives that Pt has a maximum Kerr angle of 10 nrad per A/m at 4.1 eV, while W has a maximum Kerr angle of 13 nrad per A/m at 1.2 eV[9].

A careful decomposition of our DFT calculations allow us to determine that the complex Kerr spectra at energies between 1 and 4 eV is dominated by optical transitions between d-states and s-states. Intraband transitions do not play a significant role until lower energies, e.g. less than 100 meV. As a result, the Kerr response drops rapidly at energies below the d to s interband transition threshold of 2.5 eV. Qualitatively, the Lorentzian-like response can be understood by considering how two factors affect interband transitions. At the Fermi-level, spin-accumulation leads to a change in the occupation function for up vs. down electrons, which alters the transition probabilities in Eq. (1). Spin-orbit splitting of the d-bands also affects transition probabilities. Together, these effects lead to a large polarization dependent index of refraction (Kerr rotation), and large polarization-dependent absorption (ellipticity) for photon energies near the interband transition threshold.

Even at energies well below the interband transition threshold of 2.5 eV, e.g. 1.5 eV, DFT predicts a significant real Kerr rotation as a result of d- to s-state transition effects. At low energies, e.g. 1 to 1.5 eV, the effect of the transitions from d to up s-states on the real part of $\sigma_{xy}$ is large positive. Alternatively, the effect of transitions from d- to down s-states is large negative. But these two large effects do not cancel one another perfectly, leading a non-zero real part of $\sigma_{xy}$, and a non-zero real Kerr angle.

Our density functional theory calculations include effects of spin-orbit coupling on the band-structure, but do not include effects from skew or side-jump scattering, or surfaces. Our Au films are (111) textured, and spin-orbit interactions are known to cause ≈100 meV splitting of the s-p bands near the Fermi-level at Au (111) surfaces[28]. Prior work has explained near-infrared MOKE measurements of non-magnetic metals with skew-scattering theories[5], [40]. The good agreement between theory and experiment in our study suggests surfaces, skew scattering, and side-jump scattering are probably not important processes for the MOKE spectra of Au at visible and near infrared frequencies. Definitively determining the importance of surfaces and spin-dependent electronic scattering processes would require extending measurements to lower energies in the near infrared, e.g. less than 1 eV. At such low frequencies, our theory predicts that the effect of d-state to s-state transitions is negligible, see Fig. 4. Therefore, contributions from other effects would be easier to resolve at low energies.

Our work reports the effect of spin magnetic moments on magneto-optical spectra of Au. Prior experimental and theoretical work reports the magneto-optical spectra of noble metals in the presence of static magnetic fields[4]. In static field experiments, three effects give rise to magneto-optical properties: the effect of Lorentz force on free electron motion, orbital moments, and spin moments. We point out that the optical spectra in Au we observe from spin-accumulation is different from the spectra observed due to static fields[4], especially in the near infrared. The differences in spectra for these related experiments suggest that wavelength dependent measurements of MOKE could allow independent measurements of orbital vs. spin moments in a metal. Separating spin and orbital contributions to magnetism would be useful for a variety of applications. One such application is inverse Faraday effect experiments, where circularly polarized light is believed to induce both orbital and spin moments[26].

Quantitative knowledge of the complex Kerr spectra of Au offers new opportunities for the design of optical experiments for characterizing spintronic devices. For example, low noise green laser diodes are commercially available. MOKE systems built with such green-lasers can use Au thin-films as a transducer for the measurement of spin-accumulation. Assuming a MOKE noise floor of ~100 nrad, a MOKE system with a green laser would have sensitivity to ~0.1 A/m, or ~2 spins per million atoms. Such sensitivity to small magnetic moments would enable experiments not otherwise possible[41]. Similarly, our results will aid in the design of experiments designed to study ultrafast magnetization dynamics of complex magnetic heterostructures. Prior studies have shown that separate measurements of both the real and imaginary Kerr signals allows independent determination of the time-dependent magnetization dynamics of as many as two layers[42], [43]. Scanning wavelengths to adjust sensitivity to specific layers in a heterostructures offers the opportunity to understand dynamics in more complex multilayer structures.

## IV. Conclusion

We measured the magneto-optic response of gold due to spin accumulation for photon energies in the near infrared (1.3 -1.7 eV) and visible (2.4 – 3.1 eV) spectrum. We observed a sharp transition in the Kerr rotation as a consequence of the interband transition at E = 2.50 eV. Our results display a good agreement with the DFT theoretical model, suggesting that extrinsic scattering processes do not provide a significant contribution to the magneto-optic response in gold at energies between 1.3 and 3.1 eV. Instead, the Kerr response of Au appears to be dominated by optical transitions between the s-bands and d-bands. We find that for photon energies near 2.5 eV, Au is an extremely sensitive optical magnetometer. These results provide opportunities for the development of ultra-sensitive optical spintronic devices.

## V. Experimental methods

### A. Sample preparation

We grew the metal multilayer on a (0001)-oriented $Al_2O_3$ substrate (*MTI Corporation*). The substrate was thoroughly cleaned and annealed at 1100°C in air for one hour to improve surface quality. After annealing, we deposited a 4 nm Ta/4 nm Au/(1 nm Co / 2 nm Au)×4 multilayer using an AJA Orion sputtering system, in a 3.5 mTorr Ar atmosphere with a power of 200 W. The sample stage was rotated during deposition at 20 rpm. After deposition of the Au/Co multilayer, we annealed in-situ at 250°C for 30 minutes to sharpen the multilayer interfaces. After cooling down back to room temperature, we stopped the stage rotation, and, without breaking the vacuum, we rotate a metal shutter to partially cover the substrate. The spacing between the shutter and sample was ~5 mm. We then sputter deposited a thick Au layer. The shutter shadows part of the substrate from Au deposition and causes the Au layer to have a thickness gradient (wedge layer), with a thickness between 0-300 nm.

### B.    Pump/Probe Measurements

We performed time domain thermoreflectance (TDTR) and TRMOKE experiments on our sample using a pump-probe system built around a Mai-Tai Ti:sapphire laser with a repetition rate of 80 MHz, and wavelength tunable between 700 and 1050 nm. A lithium triborate crystal serves as a second-harmonic generator to extend the wavelength range of the probe beam to 400 to 525 nm. A schematic of experimental setup is shown in Fig. 1a. Further details of our pump/probe apparatus are reported in Gomez et al[37]. The pump beam impinges on the sample through the sapphire substrate and is absorbed by the metallic multilayer sample. The probe beam reflects from the Au surface at the opposite side of the metal multilayer. The reflected probe beam is then directed towards a detection line so that pump-induced changes in the polarization of the reflected probe beam can be measured. The detection line includes a beam splitter (*Newport 10BC17MB.1* and *10BC17MB.2*), liquid crystal variable retarder (LCVR) (*Thorlabs LCC1423-A* and *LCC1423-B*), super-achromatic half wave plate (*Thorlabs SAHWP05M-700*), Wollaston prism (*Newport 10WLP08*), and balanced photodetector (*Thorlabs PDB450C*), see Fig. 1a. We carefully aligned the beam with a series of apertures to ensure either normal or 45-degree angle of incidence with all optical elements. We found even slight misalignments of a few degrees hindered our ability to accurately measure real vs. imaginary Kerr signals. This sensitivity to alignment is likely due to the dependence of the phase of the laser beam on the angle of incidence with certain optical elements, e.g., the wave-plate, optical filters, and beam-splitter.

The LCVR allows us to compensate for ellipticity introduced to the probe beam by the detection line reflecting elements (beam splitter and mirror). We calibrate the retardance of the LCVR by placing a 45° polarizer up-stream of the optical elements to be compensated (in this case, a beam splitter and mirror). We measure the polarization state of the beam immediately after the LCVR with either a visible or near-infrared polarimeter (*Thorlabs PAX 1000*). By adjusting the voltage of the LCVR, we control the net retardance caused by the detection line optical elements and LCVR. For measurements of the real part of the Kerr angle (ellipticity $\varepsilon = 0$), we set the net retardance to zero by adjusting the voltage until the probe beam is linearly polarized after the LCVR. For measurements of the imaginary part of the Kerr angle, the net retardance is set to λ/4 by adjusting the LCVR voltage until the probe beam is circularly polarized. After the proper LCVR voltages are determined for that particular laser wavelength, we remove the 45° polarizer prior to performing pump/probe measurements of the Au multilayer. The above process was repeated at each wavelength we studied, i.e. 16 times.

During pump/probe measurements, we applied an external magnetic field of ~ ±2 kOe on the sample. The 2kOe field is enough to saturate the magnetic moment of the Au/Co multilayer in the out-of-plane direction, see the MOKE measurement of the hysteresis loop in Fig. 1b. We orient the half wave plate before the Wollaston prism until there is equal intensity on both inputs of the balanced photodetector. Then, the voltage output of the balanced photodetector is proportional to the change in polarization caused by the Au. For TDTR measurements, we block one of the inputs of the balanced photodiode. Then, the voltage output by the detector is proportional to the change in reflected intensity.

## VI. Acknowledgements

The experimental work by V. O. and R. W. was supported by the U.S. Army Research Laboratory and the U.S. Army Research Office under Contract/Grant No. W911NF-18-1-0364 and W911NF-20-1-0274. S. C. acknowledges support from NSF (DMR-1848074).

## VII. Keywords

Magneto-optical effects, spin accumulation, gold thin film, ultrafast spectroscopy.

## VIII. Conflict of interest

The authors declare no conflict of interest.

## IX. Data availability

The data that support the findings of this study are available from the corresponding author upon reasonable request.

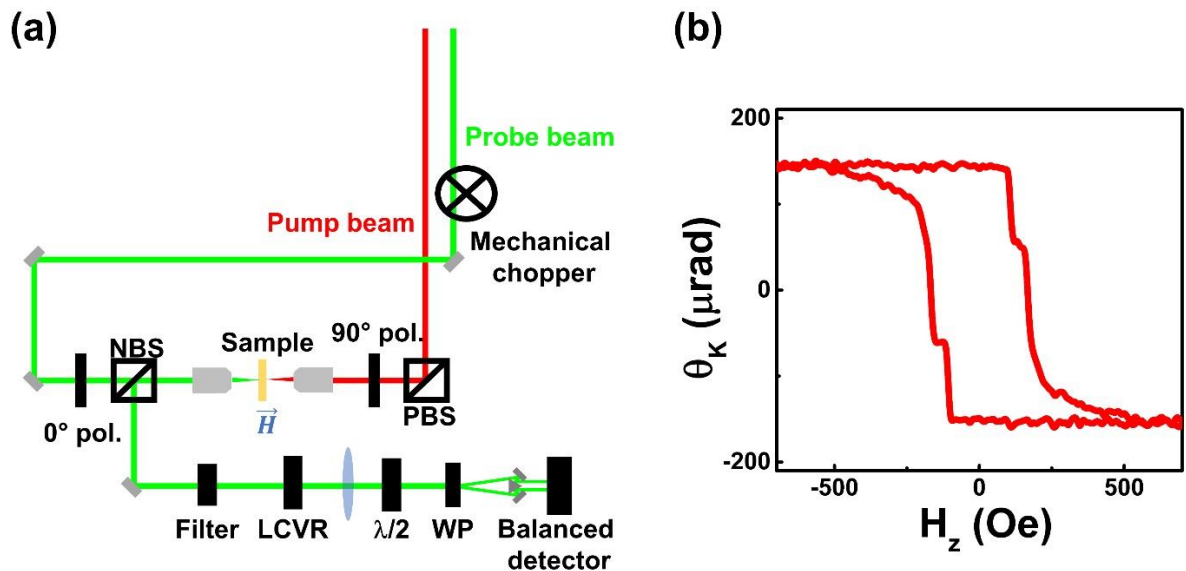

*Figure 1 a) Schematic of the pump/probe experiments. The pump beam heats up the sample, causing the spin injection into the Au layer, the reflected probe beam goes through a filter to eliminate any leaked possible pump beam, then through the liquid crystal variable retarder (LCVR), the half-wave plate (λ/2) and the Wollaston prism (WP). b) Magnetic Field-dependent polar magneto-optic Kerr effect of the Au/Co multilayer.*

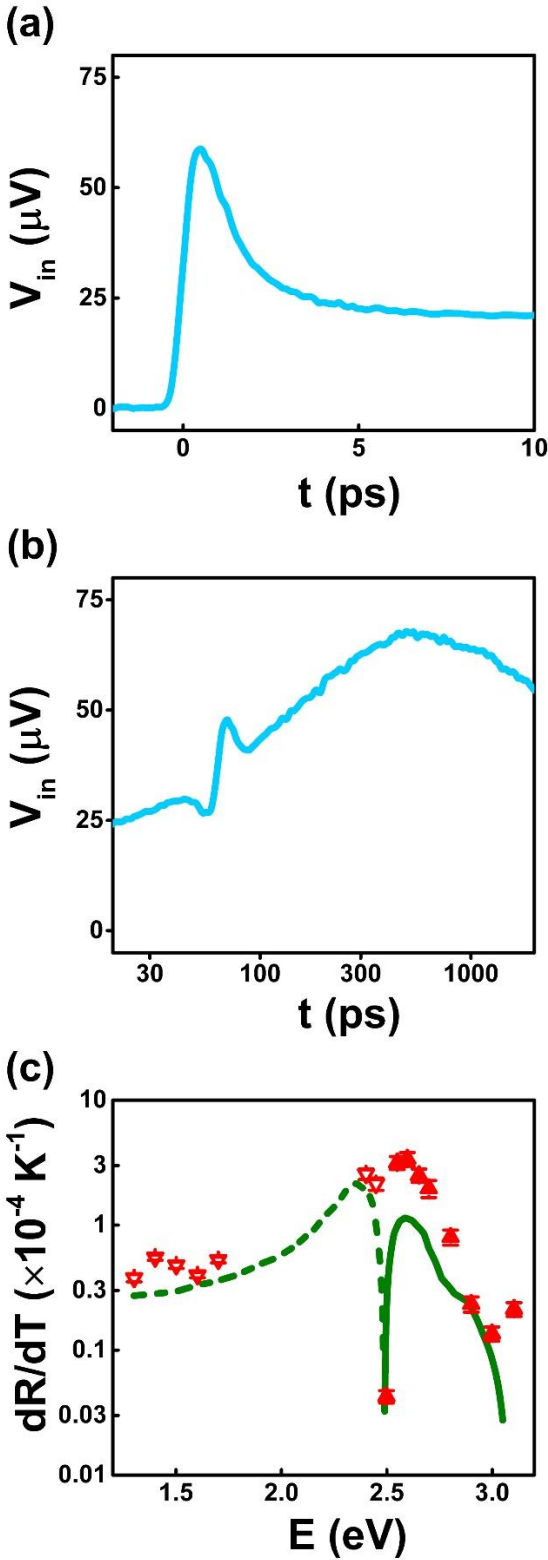

*Figure 2* Time-domain thermoreflectance measurements. a) In-phase signal vs. delay time with a probe energy E= 2.60 eV. Signals at picosecond time-scales reflect energy transport by hot electrons. b) In-phase signal vs. time delay for probe photon energy E = 2.60 eV. The feature at 61 ps is due to a longitudinal strain-wave. A 61 ps time-of-flight corresponds to a Au film thickness of 210 nm. The maximum in thermoreflectance signal occurs at 500ps, which is time

*required for the Au film to thermalize. c) Thermoreflectance vs. photon energy measured for the Au thin film used in this experiment. Thermoreflectance values were derived from probe energy dependent TDTR signals. Filled symbols correspond to positive thermoreflectance values. Open symbols correspond to negative values. The green line is experimental data for thermoreflectance spectra of Au from Wilson et al*[36].

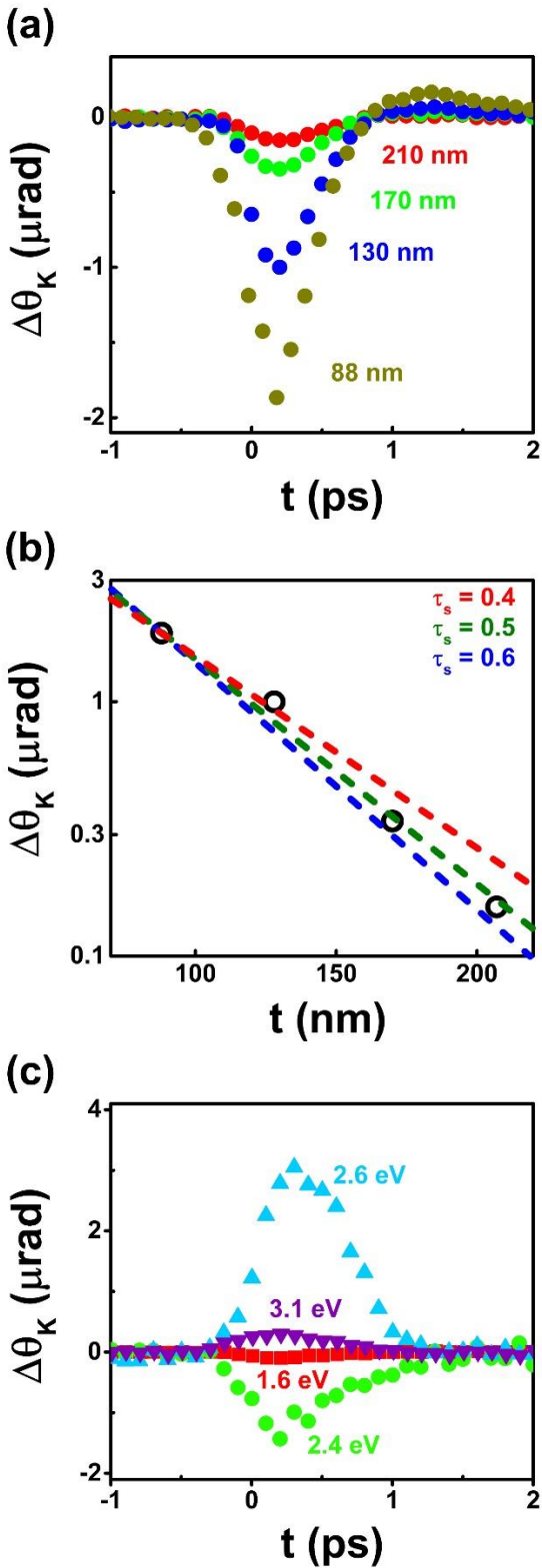

*Figure 3* Time-resolved magneto-optic Kerr effect experiments. a) Kerr rotation vs. pump/probe time delay for a photon energy E = 1.70 e for different Au film thicknesses; b) dependence of the peak Kerr rotation Kerr rotation vs. film thickness, black circles correspond to experimental

*values. Dashed lines are from calculations done in Choi et al.[5]; c) Kerr rotation of Au vs. pump/probe time delay for different photon energies.*

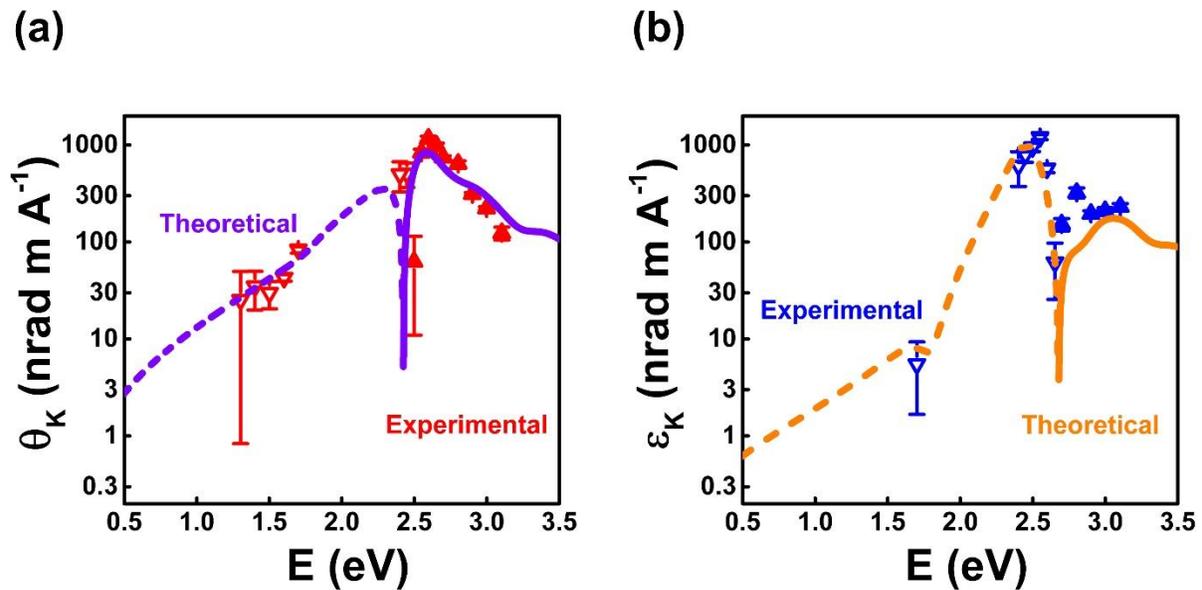

*Figure 4 Photon energy-dependent Kerr angles per magnetic moment for a Au film. a) Kerr rotation and b) Kerr ellipticity of Au. The magnetic moment is obtained by using the conversion factor obtained in Kimling et al[10]. Filled symbols correspond to positive values, while open symbols correspond to negative values. Values below the floor noise-level were omitted. The lines correspond to the theoretical predictions obtained by density functional theory, solid lines correspond to positive values, dashed lines correspond to negative values.*